\documentclass[prd,aps,showpacs,nofootinbib]{revtex4}
\usepackage{amssymb}
\usepackage{graphicx}
\usepackage{amsmath}

\newcommand{\be}{\begin{equation}}
\newcommand{\ee}{\end{equation}}
\newcommand{\bq}{\begin{eqnarray}}
\newcommand{\eq}{\end{eqnarray}}

\begin{document}

\title{Magnetic Moment Generation from non-minimal couplings in a scenario
with Lorentz-Symmetry Violation}
\author{H. Belich $^{a,e,f}$, L.P. Colatto$^{b,e}$, T. Costa-Soares$^{d,e}$,
J.A. Helay\"{e}l-Neto$^{c,e}$, M.T.D. Orlando$^{a,e}$ }
\affiliation{$^{a}${\small {Universidade Federal do Esp\'{\i}rito Santo (UFES),
Departamento de F\'{\i}sica e Qu\'{\i}mica, Av. Fernando Ferrari S/N, Vit%
\'{o}ria, ES, CEP 29060-900, Brasil}}}
\affiliation{$^{b}${\small {CEFET-Campos, Rua Dr. Siqueira 273, Campos dos Goytacazes,
RJ, CEP 28030-130, Brasil }}}
\affiliation{{\small {~}}$^{c}${\small {CBPF - Centro Brasileiro de Pesquisas F\'{\i}%
sicas, Rua Xavier Sigaud 150, Rio de Janeiro, RJ, CEP 22290-180, Brasil}}}
\affiliation{$^{d}${\small {Universidade Federal de Juiz de Fora (UFJF), Col\'{e}gio T%
\'{e}cnico Universit\'{a}rio, Av. Bernardo Mascarenhas 1283, Juiz de Fora,
MG, CEP 36080-001, Brasil}}}
\affiliation{$^{e}${\small {Grupo de F\'{\i}sica Te\'{o}rica Jos\'{e} Leite Lopes, C.P.
91933, CEP 25685-970, Petr\'{o}polis, RJ, Brasil}}}
\affiliation{$^{f}${\small {International Center for Condensed Matter Physics, Universidade
de Brasilia, Caixa postal 04667, 7910900, Brasilia-DF, Brazil}}}
\email{belich@cce.ufes.br, lcolatto@cefetcampos.br tcsoares@cbpf.br,
helayel@cbpf.br, orlando@cce.ufes.br}

\begin{abstract}
This paper deals with situations that illustrate how the violation of
Lorentz symmetry in the gauge sector may contribute to magnetic moment
generation of massive neutral particles with spin-$\frac{1}{2}$ and spin-1.
The procedure we adopt here is based on Relativistic Quantum Mechanics. We
work out the non-relativistic regime that follows from the wave equation
corresponding to a certain particle coupled to an external electromagnetic
field and a background that accounts for the Lorentz symmetry violation, and
we read thereby the magnetic dipole moment operator for the particle under
consideration.We keep track of the parameters that govern the non-minimal
electromagnetic coupling and the breaking of Lorentz symmetry in the
expressions we get for the magnetic moments in the different cases we
contemplate. Our claim is that the tiny magnetic dipole moment of truly
elementary neutral particles might signal Lorentz symmetry violation.
\end{abstract}

\pacs{11.30.Cp, 12.60.Cn,13.40.Em.}
\maketitle

\section{Introduction}

Lorentz-violating theories have been extensively studied and used as an
effective probe to test the limits of Lorentz covariance. Nowadays, these
theories are encompassed in the framework of the Extended Standard Model
(SME), \cite{Colladay},\ as a possible extension of the minimal Standard
Model of the fundamental interactions. Such kind of idea has driven much
attention mainly after some authors argued the possibility of \ Lorentz and
CPT spontaneous breaking in the context of string theory \cite{Samuel}. The
SME is the suitable framework to investigate properties of Lorentz violation
on physical systems involving photons \cite{photons1}, \cite{photons2},
radiative corrections \cite{Radiative}, fermions \cite{fermions}, neutrinos 
\cite{neutrinos}, topological defects \cite{Defects}, topological phases 
\cite{Phases}, cosmic rays \cite{CosmicRay}, supersymmetry \cite{Susy},
particle decays \cite{Iltan}, and other relevant aspects \cite{Lehnert1}, 
\cite{General}. The SME has also been used as a framework to propose Lorentz
symmetry violation \cite{Tests} and CPT\ \cite{CPT} probing experiments,
which have amounted to the imposition of stringent bounds on the
Lorentz-symmetry violating (LV) coefficients.

To take into account how this violation is implemented, in the fermion
sector of the SME, for example, there are two CPT-odd terms, $v_{\mu }%
\overline{\psi }\gamma ^{\mu }\psi ,b_{\mu }\overline{\psi }\gamma
_{5}\gamma ^{\mu }\psi $, where $v_{\mu },b_{\mu }$ are the LV backgrounds.
The modified Dirac theory has already been examined in literature \cite%
{Hamilton}, and its non-relativistic limit, with special attention to the
hydrogen spectrum \cite{Manojr} is realized. A similar study has also been
developed for the case of a non-minimal coupling with the background, with
new outcomes \cite{Nonmini}.\textbf{\ }Atomic and optical physics are other
areas in which Lorentz symmetry violation has been intensively studied.
Indeed, there are several works examining Lorentz violation in
electromagnetic cavities and optical systems \cite{Cavity}, \cite{Masers},
which contributed to establish upper bounds on the LV coefficients.

Works by Belinfante, Case, Fronsdal and Schwinger in the fifties \cite{fifth}%
, and Yang and Lee in the early sixties, \cite{Yang}, came to the general
result that charged truly-elementary particles, coupled to the
electromagnetic field, exhibit a gyromagnetic ratio given by the inverse of
its corresponding spin. For spin-half particles, described by the Dirac
field, this is a well-celebrated result. However, for higher spins, this
general result is not correct. Indeed, theoretical evidences, based on the
high-energy behavior of amplitudes and unitarity bounds \cite{Wein} and on
the dynamics of higher-spin particles propagating in electromagnetic
backgrounds as dictated by string theories, indicate that, for charged
genuinely elementary particles, the gyromagnetic ratio is always 2, no
matter what the spin of the particle is. For charged spin-1 vector bosons,
like the W-particles of the electroweak interactions, the value 2 is
reconciled by means of the Yang-Mills interactions that yield a non-minimal
(but renormalisable) coupling between the electromagnetic field-strength and
the potentials associated to the charged bosons \cite{ferrara}. Charged
spin-1 matter fields with self-interactions and topological terms have been
studied to provide a possible microscopic description for the origin of
Lorentz symmetry violation\cite{cpw}.

In our contribution, we reassess this issue in an environment dominated by a
background vector that parametrises a tiny violation of Lorentz symmetry. We
come to the conclusion that, also independently of the spin of the particle,
the Lorentz-symmetry violating background vector may yield the same
contribution to the magnetic moment and the Aharonov-Casher phase of the
particle, even if it is electrically neutral; whenever a particular
non-minimal coupling of the particle to the electromagnetic field and the
Lorentz-breaking vector is considered. Everything goes as if
Lorentz-symmetry violating background endows each elementary particle, even
those spinless and electrically neutral, with a universal contribution to
its magnetic moment and, consequently, to its Aharonov-Casher phase. In this
context, we come back to the interesting question that concerns the magnetic
properties of neutrinos\cite{jwv}. Incidentally, from Neutrino Physics, more
specifically, from the observation of non-zero neutrino masses, there
emerges a striking evidence in favour of a Physics Beyond the Standard Model%
\cite{troden}. More recently, theoretical bounds for the neutrino mass and
magnetic moment have been calculated that could be tested in the new
experiments \cite{cal}. In our considerations, we propose that magnetic
moment contributions to neutral and Majorana fermions can be obtained
already at the tree-level approximation by means of non-minimal couplings.
Besides the cases of the charged and neutral massive spin-1 particles and
the Majorana fermions themselves, another contribution we shall present in
this work refers to the way the $\left( k_{F}\right) _{\mu \nu \kappa
\lambda }$-parameter \cite{Colladay} may contribute to the magnetic dipole
moment of neutral vector bosons. This result shall be used to give us a
possible experimental bound on the magnetic moment of a neutral massive
spin-1 particle. 

One of our motivations to consider the magnetic moment contributions from
Lorentz-symmetry breaking for massive neutral particles is partly based on
the fact that this issue is always discussed in connection with (loop)
radiative corretions in field-theoretic models. Our focus is to set up a
discussion, at the level of Quantum Mechanics which, in field theory, would
correspond to the generation of magnetic moment contributions (for neutral
particles) at the tree-level. We understand that, in a scenario where the
Lorentz covariance is violated, the symmetry breaking parameters may be
responsible for the appearance of a magnetic moment for neutral particles at
the tree-level. So, we adpot this scenario to discuss magnetic properties of
neutral particles at the quantum-mechanical level. We stress that this is
one of the main goals of our work. Indeed, the discussion on magnetic dipole
moments for neutral particles is a question of relevance in connection with
results \ coming from some Physics Beyond the Standard Model. Finally, we
would like to point out that the paper of ref.\cite{gamma} reports an
interesting \ calculation of the photon magnetic moment in connection with
(external) strong magnetic fields.

The organization of our paper is given as follows: in Section II, we briefly
report on the attainment of the value 2 for the gyromagnetic ratio for
massive charged spin-1 bosons non-minimally coupled to an electromagnetic
field. In Section III, we introduce the Lorentz symmetry violation term and
we discuss the magnetic moment of a neutral massive spin-1 boson
non-minimally coupled to the background associated to the breaking of
Lorentz symmetry and an external electromagnetic field. A $\left(
k_{F}\right) -$contribution to the magnetic moment of spin-$1$ bosons is
also reported in this Section.The discussion involving Majorana fermions is
carried out in Section IV. Finally, we cast our Concluding Remarks in
Section V.

\section{Massive and Charged Spin-1 Field.}

We start off from the Lagrangian that describes a massive charged vector
matter field, $W_{\mu }$, minimally coupled to an external electromagnetic
field as below: 
\begin{equation}
\mathcal{L}=-\frac{1}{2}(W^{\mu \nu })^{\ast }W_{\mu \nu }+m^{2}(W^{\mu
})^{\ast }W_{\mu },
\end{equation}%
where $W^{\mu \nu }=D^{\mu }W^{\nu }-D^{\nu }W^{\mu },$ $D_{\mu }=\partial
_{\mu }+ieA_{\mu }$. The minimal electromagnetic coupling yields a wrong $%
(g=1)$ gyromagnetic ratio: this may be found by considering the Pauli - type
equation for the charged vector boson in the presence of a magnetic field,
namely, 
\begin{equation}
\frac{1}{2m}\left( \vec{p}-e\vec{A}\right) ^{2}W_{i}-\frac{e}{2m}\vec{B}%
\cdot \text{ }\vec{S}_{ij}W_{j}=E_{nr}W_{i},
\end{equation}%
where $E_{nr}$ means the non-relativistic energy, and $\vec{S}_{ij}$ is the
spin matrix: 
\begin{equation}
\vec{\mu}=\frac{e}{2m}\vec{S}\;,
\end{equation}%
$(S_{k})_{ij}=-i\varepsilon _{kij}$. To by-pass the conflict of the
gyromagnetic ratio, we have to introduce a renormalisable non-minimal
electromagnetic coupling \cite{ferrara}. The motivation for this new
interaction term comes from the Electroweak Theory: its $SU(2)\times U(1)$
gauge symmetry dictates the coupling between $A_{\mu }$ and the charged
gauge bosons as given below, after the spontaneous breaking of SU(2) takes
place: 
\begin{equation}
ieF_{\mu \nu }W^{\mu \ast }W^{\nu }.
\end{equation}%
This indeed cancels high-energy divergences and corrects the gyromagnetic
factor to the right value $g=2,$ as it should be. So, taking into account
this new interaction, one may consider the vector field equation that
follows: 
\begin{equation}
D_{\mu }W^{\mu \nu }+m^{2}W^{\nu }+ieW_{\mu }F^{\mu \nu }=0.
\end{equation}%
It directly implies the subsidiary condition $D_{\mu }W^{\mu }=0$. In the
non-relativistic limit, this condition yields: 
\begin{equation}
W^{0}\cong \frac{\vec{p}}{m}\cdot \vec{W}-\frac{e}{m}\vec{A}\cdot \vec{W},
\end{equation}%
which shows that the time component of the $W$-field is of \ the order $%
\displaystyle{\left( \frac{v}{c}\right) }$ of its space components. It is
worthy to remark that, by considering the time component of the field
equation above, one exactly arrives at the same relation that follows from
the subsidiary condition: we then go straight to consider the space
components of the field equation for $W^{\mu }$. By properly \ carrying out
the non-relativistic approximation, after some algebraic steps, we show that
the gyromagnetic ratio comes out with its correct value equal to $2$: 
\begin{equation}
\frac{1}{2m}\left( \vec{p}-e\vec{A}\right) ^{2}W_{i}-\frac{e}{m}\vec{S}%
_{ij}\cdot \vec{B}W_{j}=EW_{i}.\text{\ }
\end{equation}%
In the course of these calculations, it becomes clear that the net effect of
the non-minimal coupling, inherited from the non-Abelian $SU(2)$ symmetry of
the Electroweak Theory, is to add up the piece which was missing to yield
the right value for $g$. We now turn into the discussion of the gyromagnetic
ratio in situations where there occurs violation of Lorentz symmetry. This
is motivated by the fact that one may use magnetic moment measurements of
higher spin particles to get new bounds on the Lorentz symmetry violation
parameter.

\section{A Lorentz-symmetry violating background parametrised by a 4-vector, 
$v_{\protect\mu }$}

Assuming that the background vector couples to the electromagnetic field,
the covariant derivative operator can be modified to introduce a non-minimal
coupling according to the expression below: 
\begin{equation}
D_{\mu }=\partial _{\mu }+ieA_{\mu }+igv^{\alpha }\widetilde{F}_{\mu \alpha
},\text{ }
\end{equation}%
where $\widetilde{F}_{\mu \alpha }$ stands for the dual of the
electromagnetic field strength. It is worthy to mention that
Lorentz-symmetry violation does not conflict with gauge invariance. Gauge
symmetry is not violated by the action term $\varepsilon _{\mu \nu \kappa
\lambda }v^{\mu }A^{\nu }F^{\kappa \lambda }$\ introduced by Carrol, Field
and Jackiw\cite{photons1}. So, if we are to consider the dynamics of charged
particles under the action of the electromagnetic field, a gauge covariant
derivative must be adopted. In our proposal, we go a step further: we
extended the usual covariant derivative by adding up a term that implements
a (non-minimal) coupling of the particle to the external $A^{\mu }-$field
and, contemporarily, to the background vector, $v^{\mu }$. The term $%
v^{\alpha }\widetilde{F}_{\mu \alpha }$ is clearly gauge invariant, so it
does not harm the status of $D_{\mu }$ as a covariant derivative. This means
that a local phase transformation, $e^{i\alpha }$, performed on the charged
matter fields acts upon $D_{\mu }$\ according to the usual gauge
transformation of a genuine covariant derivative: $D_{\mu }^{^{\prime
}}=e^{i\alpha }D_{\mu }e^{-i\alpha }.$

In $\left( 1+2\right) D$, due to the fact that the Levi-Civita tensor is a
rank-3 tensor, the dual of the field strength is a vetor; so, we can define
a covariant derivative\cite{khare} as below:%
\begin{equation}
{\mathcal{D}}_{\mu }=\partial _{\mu }+ieA_{\mu }+ig\widetilde{F}_{\mu }.
\end{equation}%
A direct consequence of the non-minimal coupling introduced in $D_{\mu }$\
is that scalar particles display a non-trivial magnetic moment. Another
contribution of this covariant $\left( 1+2\right) D$\ derivative is the
geneneration of electrically charged vortices in the Abelian Higgs Model\cite%
{khare}. 

Based on the result referred to above, and considering that the $v^{\mu }-$\
background may, in some special case, lead to an effective $\left(
1+2\right) D$\ model $\left( v^{\mu }=\left( v^{0},v^{1},v^{2},0\right)
\right) $, we then introduce the term $igv^{\mu }\widetilde{F}_{\mu \nu }$\
as the $4-$dimensional counterpart (whenever there is Lorentz-symmetry
breaking) of the non-minimal term studied in\cite{khare}. As a consequence,
we may investigate electrically charged vortices in the $4D$\ Abelian Higgs\
model \cite{General} and the anomalous magnetic moment generation of spin-$%
\frac{1}{2}$\ particles also in $4$-dimensional space-times\cite{Phases}.

\textbf{\bigskip }The effect of the non-minimal interaction term above on a
charged vector field, as considered in the previous Section (but, now, with
the covariant derivative modified as above), is to endow the particle
associated to the $W$-field with a universal magnetic moment given by 
\begin{equation}
\vec{\mu}=\frac{1}{2}g\vec{v},  \label{mu}
\end{equation}%
as it is the case for the scalar and the spin-$\frac{1}{2}$, according to
the results reported in the work of reference \cite{Phases}. This is a very
peculiar outcome. Everything happens as if the presence of the background
modifies the structure of the particle and endows it with the universal
magnetic moment given above. According to the previous studies carried out
by Colladay and Kosteleck\'{y}, in the works of reference \cite{Colladay},
different particle species may have different independent Lorentz-breaking
parameters. Our non-minimal coupling present in $D_{\mu }$ is taken the same
for all charged particle species, for it accompanies the minimal coupling
term in the covariant derivative defined above. In the same way the minimal
coupling is universal, our term $gv^{\alpha }\widetilde{F}_{\mu \alpha }$
follows the same pattern. What is highlighted here is that its net effect,
no matter which spin the particle possesses, is to yield the same value for $%
\overrightarrow{\mu }$, as given above, in eq. \ref{mu}.

This is also the situation in the case of neutral vector particles, once the
non-minimal coupling above is switched on. Indeed, to explicitly see this
result, we take the simpler case where a non-charged ($e=0$) massive spin-$1$
particle is non-minimally coupled to the background and to the
electromagnetic field as described in the wave equation given below: 
\begin{equation}
\left( \partial _{\mu }+igv^{\kappa }\widetilde{F}_{\mu \kappa }\right)
Z^{\mu \nu }+m^{2}Z^{\nu }=0,  \label{cb1}
\end{equation}%
where $Z_{\mu }$ is the wave function of the spin-1 particle. The subsidiary
condition in this case takes the form: 
\begin{equation}
\partial _{\mu }Z^{\mu }+igv^{\kappa }\widetilde{F}_{\mu \kappa }Z^{\mu }=0,
\label{dmuzmu}
\end{equation}%
where we can notice that, differently from the Proca case, it sets up
non-trivial relations among the $Z$-field components. (Incidentally, by
introducing an external electric field, we can get how the Aharonov-Casher
(AC) phase looks like.) In the non-relativistic limit, the subsidiary
condition yields: 
\begin{equation}
Z^{0}\cong \frac{\vec{p}}{m}\cdot \vec{Z}-\frac{1}{m}(g\vec{v}\times \vec{E}%
)\cdot \vec{Z},
\end{equation}%
where we point out the presence of a sort of AC term. By replacing the
expression above for $Z^{0}$ in the space components of the $Z^{\mu }$%
-equations, and by properly keeping he terms that survive the
non-relativistic approximation, we get: 
\begin{equation}
\frac{1}{2m}\left[ \vec{p}+\frac{g}{2}\left( \vec{v}\times \vec{E}\right) %
\right] ^{2}Z_{i}=E_{nr}Z_{i}.
\end{equation}%
This result suggests \ that the quantity $\frac{1}{2}g\vec{v}$ could be
interpreted as the magnetic moment acquired by the neutral particle due to
the presence of the background vector, $\vec{v}$. We can observe that the
immediate consequence is the appearance of a universal AC phase for
different spins, by virtue of the breaking of Lorentz symmetry under the
particle point of view. Moreover, in the presence of an external electric
field, $\vec{E}$, the wave function of every particle, charged or neutral,
with or without spin, acquires a non-trivial phase given by $\frac{g}{2}%
\left( \vec{v}\times \vec{E}\right) $. To show that the quantity $\frac{1}{2}%
g\vec{v}$ is actually the magnetic moment, we consider our neutral vector
particle under the action of an external magnetic field, $\vec{B}$. To do
that, we take Eq. \eqref{cb1} and we switch on a constant magnetic field
given by 
\begin{equation}
F_{ij}=-\varepsilon _{ijk}B_{k}.
\end{equation}%
From the subsidiary condition, we get that 
\begin{equation}
Z^{0}\cong \frac{\vec{p}}{m}\cdot \vec{Z}-\frac{gv^{0}}{m}\vec{B}\cdot \vec{Z%
},
\end{equation}%
and, by means of this relation and the space components of Eq. \eqref{cb1}
taken in the non-relativistic limit, the resulting wave equation for the
space components, $Z_{i},$ reads as below: 
\begin{equation}
\frac{{\vec{p}}^{\,2}}{2m}Z_{i}+\frac{1}{2}g\vec{v}\cdot \vec{B}%
Z_{i}=E_{nr}Z_{i},
\end{equation}%
where it is easy to recognise the gyromagnetic ratio and to see that it gets
the same expression as in the scalar and spin-$\frac{1}{2}$ cases, as we had
already mentioned.

Before closing this Section and getting to the discussion on the Majorana
fermions, we belive it is worthwhile to mention another result valid for the
case of the neutral spin-1 bosons, namely, the contribution of the $k_{F}-$%
parameter\cite{Colladay} to the magnetic dipole moment of this sort of
particle, previously described by $Z^{\mu }$(present section). 

The $k_{F}-$violationg term modifies the $Z^{\mu }-$field equations as given
below:%
\begin{equation}
D_{\mu }Z^{\mu \nu }-\frac{1}{2}k_{F}^{\nu \kappa \lambda \rho }D_{\kappa
}Z_{\lambda \rho }+m^{2}Z^{\nu }=0,  \label{betha}
\end{equation}%
where 
\begin{equation}
{\mathcal{D}}_{\mu }=\partial _{\mu }+igv^{\alpha }\widetilde{F}_{\mu \alpha
}.
\end{equation}

Following along the same steps as we have shown previously (Section II), we
place the spin-$1$\ particle in an external magnetostatic field and consider
the non-relativistic regime of the corresponding field equation to read off
its corresponding magnetic moment contribution. We calculate the subsidiary
condition out of the equation above and, by considering the space components
of these field equations, where we insert the expression for $Z^{0}$ coming
from the subsidiary condition, we get, after some algebraic manipulations
and the use of the conditions for the non-relativistic regime, that the $%
k_{F}-$parameter induces the correction given by%
\begin{equation}
\overrightarrow{B}_{\cdot }\overrightarrow{\mu }_{ij}Z_{j},
\end{equation}%
where the n-th component of $\overrightarrow{\mu }_{ij}$ is given by%
\begin{equation}
\left( \mu _{n}\right) _{ij}=\frac{1}{2}gv^{0}\left( k_{F}\right) _{nij0}.
\end{equation}

With this result, in our Concluding Remarks, we shall be able to present a
bound on the $Z^{0}$ `s magnectic moment. For that, we propose a discussion
on the magnetic moment of Majorana-type neutrinos in the sequel, from which
we will be able to get information on the product of parameters $gv^{0}$. As
for $k_{F}$, we shall be adopting a result presented in the work of ref. 
\cite{klink}, so that an estimation of the magnetic moment given in the
expression above can be obtained.

\section{The case of Majorana fermions}

Neutrino magnetic dipole moments in the Standard Model are calculated as
radiative corrections and the tiny values obtained from loop calculations
may be used as good precision tests. In our case, we adopt the same
procedure followed to study the case of (massive) neutral vector bosons: we
assume a tiny deviation from the situation of Lorentz symmetry and we
non-minimally couple the (neutral) Majorana fermions to an external
electromagnetic field and the background vector that parametrizes the
breaking of the relativistic symmetry. To implement this scenario, we set up
the Dirac equation below for a Majorana spinor: 
\begin{equation}
i\gamma ^{\mu }(\partial _{\mu }+igv^{\nu }\widetilde{F}_{\mu \nu }\gamma
_{5})\Psi -m\Psi =0.  \label{Dirac1}
\end{equation}%
The introduction of the chirality matrix in the non-minimal electromagnetic
coupling is dictated by the Majorana character of the fermion we consider.
We adopt to work with the Majorana fermion by writing its wave function, $%
\Psi $, in the Majorana representation for the $\gamma $-matrices:

\begin{eqnarray}
&\gamma ^{0}=\left( 
\begin{array}{cc}
\sigma _{y} & 0 \\ 
0 & -\sigma _{y}%
\end{array}%
\right) ,\gamma ^{1}=\left( 
\begin{array}{cc}
i\sigma _{x} & 0 \\ 
0 & -i\sigma _{x}%
\end{array}%
\right) ,\gamma ^{2}=\left( 
\begin{array}{cc}
i\sigma _{z} & 0 \\ 
0 & -i\sigma _{z}%
\end{array}%
\right) ,\gamma ^{3}=\left( 
\begin{array}{cc}
0 & i1 \\ 
i1 & 0%
\end{array}%
\right) &,  \notag \\
&\gamma ^{5}=i\gamma ^{0}\gamma ^{1}\gamma ^{2}\gamma ^{3}=\left( 
\begin{array}{cc}
0 & -i1 \\ 
i1 & 0%
\end{array}%
\right) .&
\end{eqnarray}%
The charge conjugation matrix is $C=-\gamma ^{0}$; thus, in this
representation, a Majorana spinor $\left( \Psi ^{c}=C\bar{\Psi}^{t}=\Psi
\right) $ exhibits 4 real components. We should remark that parity may be
preserved despite the appearance of the chirality matrix in the action term $%
\bar{\Psi}\gamma ^{\mu }\gamma _{5}\Psi v^{\nu }\widetilde{F}_{\mu \nu }$ ;
the particular property of $v^{\mu }$\ under parity (vector or a
pseudo-vector) may be suitably chosen so as to make this term
parity-preserving.

As first step, we must probe the neutral particle by subjecting it to an
external magnetic field, $\widetilde{F}_{0i}=\vec{B}$. This shall reveal the
(eventual) contributions of the Lorentz-symmetry violating parameters, v0
and $\vec{v}$, to the magnetic moment of the neutral fermion. We start of
from the coupled Dirac's Eq. \eqref{Dirac1} ,

\begin{equation}
\left( \gamma ^{0}E-\vec{\gamma}\vec{p}-m-g\vec{v}\cdot \vec{B}\gamma
^{0}\gamma _{5}+gv^{0}\vec{B}\cdot \vec{\gamma}\gamma _{5}\right) \Psi =0
\end{equation}

The coupled Dirac%
\'{}%
s equation equation above splits up into 2 equations for the 2-component
spinors, $\xi $ and$\ \chi $. They read as follows:

\begin{eqnarray}
M\xi +N\chi &=&0,  \notag \\
P\xi -Q\chi &=&0
\end{eqnarray}

where

\begin{eqnarray}
&M\equiv m-E\sigma _{y}+ip_{x}\sigma _{x}+ip_{y}\sigma
_{z}+gv^{0}B_{z};\;\;N\equiv ip_{z}-ig\vec{v}\cdot \vec{B}\sigma
_{y}-gv^{0}B_{x}\sigma _{x}-gv^{0}B_{y}\sigma _{z},&  \notag \\
&P\equiv -ip_{z}+ig\vec{v}\cdot \vec{B}\sigma _{y}+gv^{0}B_{x}\sigma
_{x}+gv^{0}B_{y}\sigma _{z};\;\;Q\equiv m-E\sigma _{y}+ip_{x}\sigma
_{x}+ip_{y}\sigma _{z}+gv^{0}B_{z};&
\end{eqnarray}

\begin{eqnarray}
\chi &=&Q^{-1}P\xi ;  \notag \\
\left( M+NQ^{-1}P\right) \xi &=&0.
\end{eqnarray}

Using the quaternionic unities \cite{qua}, we cast the operators $M,N,P$ an $%
Q$ in the form below:

\begin{eqnarray}
M &\equiv &\left( m+gv^{0}B_{z}\right) +p_{x}I+iEJ+p_{y}K;  \notag \\
N &\equiv &ip_{z}+igv^{0}B_{x}I-g\vec{v}\cdot \vec{B}J+igv^{0}B_{y}K,  \notag
\\
P &\equiv &-ip_{z}-igv^{0}B_{x}I+g\vec{v}\cdot \vec{B}J-igv^{0}B_{y}K; 
\notag \\
Q &=&\left( m+gv^{0}B_{z}\right) +p_{x}I-iEJ+p_{y}K.
\end{eqnarray}%
The $(M+NQ^{-1}P)$-operator, that yields a Pauli-type equation, is worked
out, but, by analyzing its explicit form, the magnetic moment cannot be
properly identified. This shows us that the Majorana representation is not
suitable for the sake of taking the non-relativistic approximation. We
better go over into the (usual) Dirac 's representation and we also propose
a more general situation, where we try to compare the competitive effect of
two non-minimal couplings that may be contemporarily present in the Dirac%
\'{}%
s equation for a Majorana fermion; both the couplings are collected in the
expression below:%
\begin{equation}
i\gamma ^{\mu }(\partial _{\mu }+igv^{\nu }\widetilde{F}_{\mu \nu }\gamma
_{5})\Psi +\widetilde{g}F_{\mu \nu }\Sigma ^{\mu \nu }\gamma _{5}\Psi -m\Psi
=0.
\end{equation}%
So, from this complete equation, we follow the necessary steps to work out
the non-relativistic approximation and to arrive at a Pauli-type equation
for the $\xi $-component. By properly treating the terms that dominate in
the non-relativistic limit and taking care of the relations imposed by the
Clifford algebra, we find out the non-relativistic equation for $\xi $,
which turns out to be: 
\begin{equation}
i\hbar \frac{\partial }{\partial t}\xi =\left\{ \frac{1}{2m}\left( \vec{p}+2~%
\widetilde{g}~\vec{\sigma}\times \vec{B}\right) ^{2}+~g~v^{0}\vec{\sigma}%
\cdot \vec{B}-~g~\vec{E}\cdot \left( \vec{v}\times \vec{\sigma}\right) -~%
\tilde{g}~\vec{\sigma}\cdot \vec{E}\right\} \xi .
\end{equation}%
The expression above opens up a number of interesting remarks on the
(non-minimal) electromagnetic effects of the spin of a neutral
self-conjugated fermion. We identify the interaction term that leads to the
magnetic dipole moment as being given by $g~v^{0}\vec{\sigma}\cdot \vec{B}$;
this then shows that, instead of the space component, $\vec{v}$, it is now
the time component, $v^{0}$, the responsible for the magnetic moment
generation, and the Pauli-type coupling (the one given by $\widetilde{g}$)
does not contribute to the magnetic interaction as in the ordinary case.
Instead, it induces a coupling to the electric field and a new type of phase
($\Phi $) in the fermion wave function, given by 
\begin{equation}
\Phi =\frac{2~\widetilde{g}}{~g~v^{0}}\int \dot{d}\vec{l}\cdot (\vec{\mu}%
\times \vec{B}).
\end{equation}

\section{Concluding Remarks}

The main effort in our work has been to show how, in an environment where
Lorentz symmetry is violated, truly elementary neutral particles may show up
magnetic properties only due to their spin, once non-minimal couplings to
the electromagnetic field are allowed. The background vector responsible for
the Lorentz-symmetry violation couples to both the electromagnetic field and
the particle itself and then the electromagnetic properties of the spin are
revealed through Aharonov-Casher and Pauli-type couplings of the magnetic
dipole moment of the particle. For charged spin-0, spin-$\frac{1}{2}$,
spin-1 particles and neutral vector bosons, we have seen that there appears
a universal magnetic dipole moment for each particle, $\vec{\mu}=\frac{1}{2}g%
\vec{v}$, as a result of the presence of the background vector.
Nevertheless, other contributions for the magnetic moment may be derived
which depend on the type of the particle, as discussed in the treatment of
the Majorana fermion, for which the non-minimal coupling with the presence
of the chirality matrix produced a new sort of phase which involves the
magnetic field. This means that, for neutral particles like the neutrinos,
the tiny magnetic dipole moments they have (less than $10^{-10}$ Bohr
magnetons), which in the framework of the Electroweak Theory are understood
and computed as an effect of the radiative corrections, may also be
attributed to possible effects of an eventual violation of Lorentz symmetry.
What we conclude is that purely electromagnetic effects of the spin may
emerge if neutral particles interact with an external electromagnetic field
via a background that realizes the tiny breaking of Lorentz symmetry. In the
present paper, this has been done for a background vector; however, from our
results, we can safely state that the same conclusions hold through if the
Lorentz-symmetry breaking background is of a tensor nature.\ 

With the result on the magnetic moment for Majorana-type fermions presented
in the previous Section, and the experimental bounds on the neutrino
magnetic moment\cite{pdg}, we can set a bound on the product $gv^{0}$,
namely, $gv^{0}<0.9\times 10^{-10}\mu _{B}$. By considering the results of
the work of ref. \cite{rei}, and a recent paper by Klinkhamer and Shereck 
\cite{klink}, it is reasonable to take the bound $\left\vert \left(
k_{F}\right) _{nij0}\right\vert <2\times 10^{-7}$, so that we can get an
estimation on the magnetic moment for $Z^{0}$: $\mu (Z^{0})<1,65\times
10^{-14}\mu _{N}$, where $\mu _{N}$\ stands for the nuclear magneton. In
possess of the results presented in this work, we are now concentrating our
efforts to systematically get bounds on Lorentz-symmetry breaking parameters
from our investigation of their influence on the calculation of gyromagnetic
ratios and magnetic moments for different particle species, with special
interest on the sector of neutral fundamental fermions and vector bosons.
These results shall be soon reported in a forthcoming paper.

\acknowledgements

J. Moraes and R. Turcati are kindly acknowledged for long discussion. HB,
TCS, JAHN and MDTO acknowledge CNPq for the invaluable financial help.

\end{document}